\documentclass[italian,english]{article}
\usepackage[latin9]{inputenc}
\usepackage{color}
\usepackage{graphicx}
\usepackage{amssymb}

\newcommand{\lyxaddress}[1]{
\par {\raggedright #1
\vspace{1.4em}
\noindent\par}
}

\usepackage{babel}

\begin{document}

\title{\textbf{A clarification on a common misconception about interferometric
detectors of gravitational waves }}

\author{\textbf{Christian Corda}}

\maketitle

\lyxaddress{\begin{center}
International Institute for Theoretical Physics and Mathematics Einstein-Galilei,
Via Santa Gonda, 14, 59100 Prato, Italy and Institute for Basic Research,
P. O. Box 1577, Palm Harbor, FL 34682, USA %
\footnote{\begin{quotation}
\emph{Partially supported by a Research Grant of the R. M. Santilli
Foundation Number RMS-TH-5735A2310}
\end{quotation}
}
\par\end{center}}

\lyxaddress{\begin{center}
\textit{E-mail address:} \textcolor{blue}{corda.galilei@gmail.com} 
\par\end{center}}
\begin{abstract}
The aims of this letter are two. First, to show the angular gauge-invariance
on the response of interferometers to gravitational waves (GWs). In
this process, after resuming for completeness results on the Transverse-Traceless
(TT) gauge, where, in general, the theoretical computations on GWs
are performed, we analyse the gauge of the local observer, which represents
the gauge of a laboratory environment on Earth. The gauge-invariance
between the two gauges is shown in its full angular and frequency
dependences. In previous works in the literature this gauge-invariance
was shown only in the low frequencies approximation or in the simplest
geometry of the interferometer with respect to the propagating GW
(i.e. both of the arms of the interferometer are perpendicular to
the propagating GW). 

Second, as far as the computation of the response functions in the
gauge of the local observer is concerned, a common misconception about
interferometers is also clarified. Such a misconception purports that,
as the wavelength of laser light and the length of an interferometer's
arm are both stretched by a GW, no effect should be visible, invoking
an analogy with cosmological redshift in an expanding universe. 
\end{abstract}
The scientific community aims in a first direct detection of GWs in
next years (for the current status of GWs interferometers see \cite{key-1})
confirming the indirect, Nobel Prize Winner, proof of Hulse and Taylor
\cite{key-2}. 

Detectors for GWs will be important for a better knowledge of the
Universe and either to confirm or to rule out, in an ultimate way,
the physical consistency of General Relativity, eventually becoming
an observable endorsing of Extended Theories of Gravity, see \cite{key-3}
for details.

In the framework of General Relativity, computations on GWs are usually
performed in the so-called TT gauge \cite{key-3,key-4}. This kind
of gauge is historically called Transverse-Traceless, because in these
particular coordinates GWs have a transverse effect, are traceless
and the computations are, in general, simpler \cite{key-4}. As interferometers
work in a laboratory environment on Earth, the gauge in which the
space-time is locally flat and the distance between any two points
is given simply by the difference in their coordinates in the sense
of Newtonian physics has to be used \cite{key-4}. In this gauge,
called the gauge of the local observer \cite{key-4,key-5,key-6},
GWs manifest themselves by exerting tidal forces on the masses (the
mirrors and the beam-splitter in the case of an interferometer, see
Figure 1). At this point, when approaching the first direct detection
of GWs, it is very important, whatever the frequency and the direction
of propagation of the GW will be, to demonstrate that the signal in
the TT gauge, which has been computed only in theoretical approaches
in the literature, is equal to the one computed in the gauge of the
local observer (which is the gauge where the detection will be observed
on Earth).

In this letter, such a gauge-invariance on the response of interferometers
to GWs between the two mentioned gauges is shown. In this process,
after a resume of results on the TT gauge which have been obtained
in \cite{key-3}, which is due for completeness, the response functions
of interferometers are computed directly in the gauge of the local
observer, obtaining the same result of the computation in the TT gauge.
In this way, the gauge-invariance is shown in its full angular and
frequency dependences. In previous works in the literature, this gauge-invariance
was shown only in the low frequencies approximation (i.e. wavelength
of the GW much large than the linear distance between test masses,
see \cite{key-7,key-8} for example) or in the simplest geometry of
the interferometer in respect to the propagating GW (i.e. both of
the arms of the interferometer are perpendicular to the propagating
GW) \cite{key-9}. 

Of course, we do not have any idea concerning the direction of the
propagating GW that will arrive to detectors, exactly like we do not
have any idea concerning its frequency. Thus, a generalization which
will take into account both of the full angular and frequency dependences
is due. The presented results are consistent with previous approximations.
As far as the computation of the response functions in the gauge of
the local observer is concerned, a common misconception about interferometers
is also clarified. Such a misconception purports that, as the wavelength
of laser light and the length of an interferometer\textquoteright{}s
arm are both stretched by a GW, no effect should be visible, invoking
an analogy with cosmological redshift in an expanding universe \cite{key-17}. 

This issue has been raised in some papers in the literature \cite{key-17}-\cite{key-20}.

In \cite{key-18} a qualitative intuitive explanation of the issue
has been discussed. In \cite{key-19} the issue has been addressed
by recalling that this is the most common question asked about interferometric
detectors of GWs. The author provided a qualitative answer \cite{key-19}:
\textquotedblleft{}\emph{Does the wavelength of the light in the gravitational
wave get stretched and squeezed the same manner as these mirrors move
back and forth? ... The answer is no, the spacetime curvature influences
the light in a different manner that it influences the mirror separations
... the influence on the light is negligible and it is only the mirrors
that get moved back and forth and the light\textquoteright{}s wavelength
does not get changed at all ...}\textquotedblright{}. In \cite{key-20}
the issue has been raised again, by invoking the analogy with the
cosmological situation. An analogy between the gauge freedom of General
Relativity and the Aharonov-Bohm effect in quantum mechanics has been
discussed too \cite{key-20}. In both situations gauge-dependent quantities
appear in the equations describing the physics but the final physical
results calculated are gauge-independent \cite{key-20}. Again, the
answer is that at the end the only physical quantity measured (the
phase shift between two laser beams in an interferometric detector)
is gauge independent. Finally, in \cite{key-17}, a direct calculation
was performed, but with the assumption of wavelength of the GW much
large than the linear distance between the beam splitter and the mirror
of the interferometer (i.e. within the low-frequency approximation).
Here the issue is ultimately clarified with a new direct calculation
and without any approximation.

Following \cite{key-3}, we work with $G=1$, $c=1$ and $\hbar=1$
and we call $h_{+}(t+z)$ and $h_{\times}(t+z)$ the weak perturbations
due to the $+$ and the $\times$ polarizations which are expressed
in terms of synchronous coordinates in the TT gauge. In this way,
the most general GW propagating in the $z$ direction can be written
in terms of plane monochromatic waves \cite{key-3,key-4} \begin{equation}
\begin{array}{c}
h_{\mu\nu}(t+z)=h_{+}(t+z)e_{\mu\nu}^{(+)}+h_{\times}(t+z)e_{\mu\nu}^{(\times)}=\\
\\=h_{+0}\exp i\omega(t+z)e_{\mu\nu}^{(+)}+h_{\times0}\exp i\omega(t+z)e_{\mu\nu}^{(\times)},\end{array}\label{eq: onda generale}\end{equation}

and the correspondent line element will be

\begin{equation}
ds^{2}=dt^{2}-dz^{2}-(1+h_{+})dx^{2}-(1-h_{+})dy^{2}-2h_{\times}dxdx.\label{eq: metrica TT totale}\end{equation}

The wordlines $x,y,z=const.$ are timelike geodesics, representing
the histories of free test masses \cite{key-3,key-4}, that, in our
case, are the beam-splitter and the mirrors of an interferometer,
see Figure 1 and ref. \cite{key-6}. 

\begin{figure}
\includegraphics{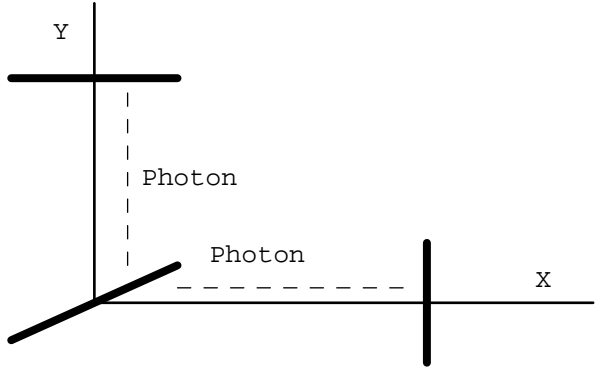}

\caption{photons can be launched from the beam-splitter to be bounced back
by the mirror, adapted from ref. \cite{key-6}}

\end{figure}

In order to obtain the response functions in the TT gauge, a generalization
of the analysis in \cite{key-9} has been be used in \cite{key-3},
the so called {}``bouncing photon method'': a photon can be launched
from the beam-splitter to be bounced back by the mirror (Figure 1).
This method has been generalized to scalar waves, angular dependences
and massive modes of GWs in \cite{key-3,key-6,key-10}. This includes
the more general problem of finding the null geodesics of light in
the presence of a weak GW \cite{key-11,key-12,key-13,key-14}. 

Defining \cite{key-3}

\begin{equation}
\begin{array}{c}
\tilde{H}_{u}(\omega,\theta,\phi)\equiv\frac{-1+\exp(2i\omega L)}{2i\omega(1+\sin^{2}\theta\cos^{2}\phi)}+\\
\\+\frac{-\sin\theta\cos\phi((1+\exp(2i\omega L)-2\exp i\omega L(1-\sin\theta\cos\phi)))}{2i\omega(1+\sin^{2}\cos^{2}\phi)}\end{array}\label{eq: fefinizione Hu}\end{equation}

and 

\begin{equation}
\begin{array}{c}
\tilde{H}_{v}(\omega,\theta,\phi)\equiv\frac{-1+\exp(2i\omega L)}{2i\omega(1+\sin^{2}\theta\sin^{2}\phi)}+\\
\\+\frac{-\sin\theta\sin\phi((1+\exp(2i\omega L)-2\exp i\omega L(1-\sin\theta\sin\phi)))}{2i\omega(1+\sin^{2}\theta\sin^{2}\phi)},\end{array}\label{eq: fefinizione Hv}\end{equation}

the total frequency and angular dependent response function (i.e.
the detector pattern) of an interferometer to the $+$ polarization
of the GW has been obtained in \cite{key-3} in the TT gauge:

\begin{equation}
\begin{array}{c}
\tilde{H}^{+}(\omega)\equiv\Upsilon_{u}^{+}(\omega)-\Upsilon_{v}^{+}(\omega)=\\
\\=\frac{(\cos^{2}\theta\cos^{2}\phi-\sin^{2}\phi)}{2L}\tilde{H}_{u}(\omega,\theta,\phi)+\\
\\-\frac{(\cos^{2}\theta\sin^{2}\phi-\cos^{2}\phi)}{2L}\tilde{H}_{v}(\omega,\theta,\phi)\end{array}\label{eq: risposta totale Virgo +}\end{equation}

that, in the low frequencies limit ($\omega\rightarrow0$) gives the
well known low frequency response function of \cite{key-15,key-16}
for the $+$ polarization: 

\begin{equation}
\tilde{H}^{+}(\omega\rightarrow0)=\frac{1}{2}(1+\cos^{2}\theta)\cos2\phi.\label{eq: risposta totale approssimata}\end{equation}

The same analysis works for the $\times$ polarization, see \cite{key-3}
for details. In that case, one obtains that the total frequency and
angular dependent response function of an interferometer to the $\times$
polarization is:

\begin{equation}
\tilde{H}^{\times}(\omega)=\frac{-\cos\theta\cos\phi\sin\phi}{L}[\tilde{H}_{u}(\omega,\theta,\phi)+\tilde{H}_{v}(\omega,\theta,\phi)],\label{eq: risposta totale Virgo per}\end{equation}

that, in the low frequencies limit ($\omega\rightarrow0$), gives
the low frequency response function of \cite{key-15,key-16} for the
$\times$ polarization: \begin{equation}
\tilde{H}^{\times}(\omega\rightarrow0)=-\cos\theta\sin2\phi.\label{eq: risposta totale approssimata 2}\end{equation}

The importance of these results is due to the fact that in this case
the limit where the wavelength is shorter than the length between
the splitter mirror and test masses is calculated. The signal drops
off the regime, while the calculation agrees with previous calculations
for longer wavelengths \cite{key-15,key-16}. The contribution is
important especially in the high-frequency portion of the sensitivity
band.

Now, let us see the computation in the gauge of the local observer.

A detailed analysis of the gauge of the local observer is given in
Ref. \cite{key-4}, Sect. 13.6. Here, we only recall that  the effect
of GWs on test masses is described by the equation for geodesic deviation
in this gauge

\begin{equation}
\ddot{x^{i}}=-\widetilde{R}_{0k0}^{i}x^{k},\label{eq: deviazione geodetiche}\end{equation}
where $\widetilde{R}_{0k0}^{i}$ are the components of the linearized
Riemann tensor \cite{key-4}. 

In the computation of the response functions in this gauge, a common
misconception about interferometers will be also clarified. Once again,
let us recall this issue. The famous misconception purports that,
as the wavelength of the laser light and the length of an interferometer's
arm are both stretched by a GW, no effect should be present, invoking
an analogy with the cosmological redshift of the expanding Universe.
This misconception has been recently clarified in a good way in \cite{key-17},
but only in the low frequency approximation. Here the misconception
will be clarified in the full angular and frequency dependences of
a GW, showing that the variation of proper time due to the photons
redshift is different from the variation of proper time due to the
motion of the arms.

We start with the $+$ polarization. In the gauge of the local observer
the equation of motion for the test masses are \cite{key-4,key-9}

\begin{equation}
\ddot{x}=\frac{1}{2}\ddot{h}_{+}x,\label{eq: accelerazione mareale lungo x}\end{equation}

\begin{equation}
\ddot{y}=-\frac{1}{2}\ddot{h}_{+}y,\label{eq: accelerazione mareale lungo y}\end{equation}

\begin{equation}
\ddot{z}=0,\label{eq: accelerazione mareale lungo z}\end{equation}

which can be solved using the perturbation method \cite{key-4}, obtaining

\begin{equation}
\begin{array}{c}
x(t)=l_{1}+\frac{1}{2}[l_{1}h_{+}(t)-l_{2}h_{\times}(t)]\\
\\y(t)=l_{2}-\frac{1}{2}[l_{2}h_{+}(t)+l_{1}h_{\times}(t)]\\
\\z(t)=l_{3},\end{array}\label{eq: traditional}\end{equation}

where $l_{1},\textrm{ }l_{2}\textrm{\textrm{ } }and\textrm{ }l_{3}$
are the coordinates of the mirror of the interferometer in absence
of GWs, with the beam-splitter located in the origin of the coordinate
system. But the arms of the interferometer are in the $\overrightarrow{u}$
and $\overrightarrow{v}$ directions, while the $x,y,z$ frame is
the proper frame of the propagating GW. Then, a spatial rotation of
the coordinate system has to be performed:

\begin{equation}
\begin{array}{ccc}
u & = & -x\cos\theta\cos\phi+y\sin\phi+z\sin\theta\cos\phi\\
\\v & = & -x\cos\theta\sin\phi-y\cos\phi+z\sin\theta\sin\phi\\
\\w & = & x\sin\theta+z\cos\theta,\end{array}\label{eq: rotazione}\end{equation}

In this way, the GW is propagating from an arbitrary direction $\overrightarrow{r}$
to the interferometer (see Figure 2 and ref. \cite{key-10}). 

\begin{figure}
\includegraphics{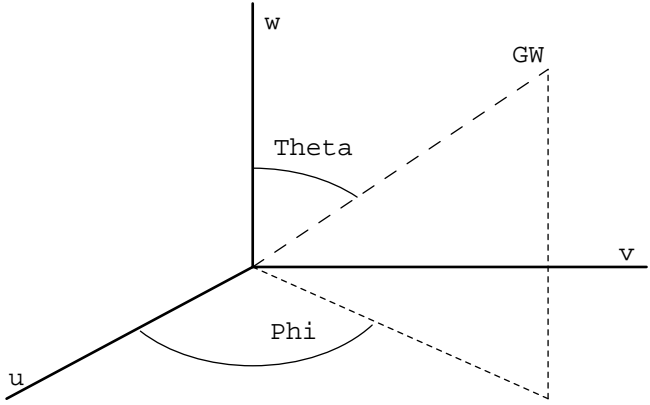}

\caption{a GW propagating from an arbitrary direction, adapted from ref. \cite{key-10}}

\end{figure}
As a result, the $u$ coordinate of the mirror is \begin{equation}
u=L+\frac{1}{2}LAh_{+}(t+u\sin\theta\cos\phi),\label{eq: du}\end{equation}

where \begin{equation}
A\equiv\cos^{2}\theta\cos^{2}\phi-\sin^{2}\phi,\label{eq: A}\end{equation}

and $L=\sqrt{l_{1}^{2}+l_{2}^{2}+l_{3}^{2}}$ is the length of the
interferometer arms.

We consider a photon which propagates in the $u$ axis. The unperturbed
(i.e. in absence of GWs) propagation time between the two masses is

\begin{equation}
T=L.\label{eq: tempo imperturbato}\end{equation}

From eq. (\ref{eq: du}), the displacements of the two masses under
the influence of the GW are

\begin{equation}
\delta u_{b}(t)=0\label{eq: spostamento beam-splitter}\end{equation}

and

\begin{equation}
\delta u_{m}(t)=\frac{1}{2}LAh_{+}(t+L\sin\theta\cos\phi).\label{eq: spostamento mirror}\end{equation}

In this way, the relative displacement, which is defined by

\begin{equation}
\delta L(t)=\delta u_{m}(t)-\delta u_{b}(t),\label{eq: spostamento relativo}\end{equation}

gives

\begin{equation}
\frac{\delta T(t)}{T}=\frac{\delta L(t)}{L}=\frac{1}{2}LAh_{+}(t+L\sin\theta\cos\phi).\label{eq: strain magnetico}\end{equation}
But, for a large separation between the test masses (in the case of
Virgo the distance between the beam-splitter and the mirror is three
kms, four in the case of LIGO), the definition (\ref{eq: spostamento relativo})
for relative displacements becomes unphysical because the two test
masses are taken at the same time and therefore cannot be in a casual
connection \cite{key-6,key-9,key-10}. The correct definitions for
the bouncing photon are

\begin{equation}
\delta L_{1}(t)=\delta u_{m}(t)-\delta u_{b}(t-T_{1})\label{eq: corretto spostamento B.S. e M.}\end{equation}

and

\begin{equation}
\delta L_{2}(t)=\delta u_{m}(t-T_{2})-\delta u_{b}(t),\label{eq: corretto spostamento B.S. e M. 2}\end{equation}

where $T_{1}$ and $T_{2}$ are the photon propagation times for the
forward and return trip correspondingly. According to the new definitions,
the displacement of one test mass is compared to the displacement
of the other at a later time, in order to allow a finite delay for
the light propagation. The propagation times $T_{1}$ and $T_{2}$
in eqs. (\ref{eq: corretto spostamento B.S. e M.}) and (\ref{eq: corretto spostamento B.S. e M. 2})
can be replaced with the nominal value $T$ because the test mass
displacements are already first order in the GW's field \cite{key-6,key-9,key-10}.
Thus, the total change in the distance between the beam splitter and
the mirror, in one round-trip of the photon, is

\begin{equation}
\delta L_{r.t.}(t)=\delta L_{1}(t-T)+\delta L_{2}(t)=2\delta u_{m}(t-T)-\delta u_{b}(t)-\delta u_{b}(t-2T),\label{eq: variazione distanza propria}\end{equation}

and, in terms of the amplitude of the $+$ polarization of the GW:

\begin{equation}
\delta L_{r.t.}(t)=LAh_{+}(t+L\sin\theta\cos\phi-L).\label{eq: variazione distanza propria 2}\end{equation}
The change in distance (\ref{eq: variazione distanza propria 2})
leads to changes in the round-trip time for photons propagating between
the beam-splitter and the mirror:

\begin{equation}
\frac{\delta_{1}T(t)}{T}=Ah_{+}(t+L\sin\theta\cos\phi+L).\label{eq: variazione tempo proprio 1}\end{equation}

In the last calculation (variations in the photon round-trip time
which come from the motion of the test masses induced by the GW),
it has been implicitly assumed that the propagation of the photon
between the beam-splitter and the mirror of the interferometer is
uniform as if it were moving in a flat space-time. But the presence
of the tidal forces indicates that the space-time is curved \cite{key-4,key-6,key-9,key-10}.
As a result, one more effect after the first discussed has to be considered,
and it requires spacial separation (note: in \cite{key-9} the effects
considered were three, but the third effect vanishes putting the beam
splitter in the origin of the coordinate system \cite{key-6}). This
is exactly the contribution of the photons redshift. If it results
different from the contribution of the test masses motion in previous
analysis (i.e. the sum of the two contributions is different from
zero), it also clarifies the misconception purporting that, because
the wavelength of the laser light and the length of an interferometer's
arm are both stretched by a GW, no effect should be present.

From equations (\ref{eq: accelerazione mareale lungo x}), (\ref{eq: accelerazione mareale lungo y})
and (\ref{eq: rotazione}) the tidal acceleration of a test mass caused
by the GW in the $u$ direction is\begin{equation}
\ddot{u}(t+u\sin\theta\cos\phi)=\frac{1}{2}LA\ddot{h}_{+}(t+u\sin\theta\cos\phi).\label{eq: acc}\end{equation}

Equivalently, one can say that there is a gravitational potential
\cite{key-4,key-6,key-9}:

\begin{equation}
V(u,t)=-\frac{1}{2}LA\int_{0}^{u}\ddot{h}_{+}(t+l\sin\theta\cos\phi)dl,\label{eq:potenziale in gauge Lorentziana}\end{equation}

which generates the tidal forces, and that the motion of the test
mass is governed by the Newtonian equation

\begin{equation}
\ddot{\overrightarrow{r}}=-\bigtriangledown V.\label{eq: Newtoniana}\end{equation}

In the framework of weak-field gravity, the interval in the gauge
of the local observer is given by \cite{key-4,key-5}\begin{equation}
ds^{2}=g_{00}dt^{2}+du^{2}+dv^{2}+dw^{2}.\label{eq: metrica osservatore locale}\end{equation}
Equations like eq. (\ref{eq: acc}) work for the $v$ an the $w$
directions too. Thus, photon momentum in these directions is not conserved
and photons launched in the $u$ axis will deflect out of this axis.
But here this effect can be neglected, because the photon deflections
into the $v$ and $w$ directions will be at most of order $h_{+}$
\cite{key-4,key-6,key-9}. Then, to first order in $h_{+}$, the $dv^{2}$
and $dw^{2}$ terms can be neglected. Thus, the line element (\ref{eq: metrica osservatore locale}),
for photons propagating along the $u$ - axis, can be rewritten as

\begin{equation}
ds^{2}=g_{00}dt^{2}+du^{2}.\label{eq: metrica osservatore locale in u}\end{equation}

The condition for a null trajectory ($ds=0$) and the well known relation
between Newtonian theory and linearized gravity ($g_{00}=1+2V$ \cite{key-4,key-5})
give the coordinate velocity of the photons 

\begin{equation}
v_{c}^{2}\equiv(\frac{du}{dt})^{2}=1+2V(t,u),\label{eq: velocita' fotone in gauge locale}\end{equation}

which, to first order in $h_{+},$ is approximated by

\begin{equation}
v_{c}\approx\pm[1+V(t,u)],\label{eq: velocita fotone in gauge locale 2}\end{equation}

with $+$ and $-$ for the forward and return trip respectively. Knowing
the coordinate velocity of the photon, the propagation time for its
travelling between the beam-splitter and the mirror can be defined:

\begin{equation}
T_{1}(t)=\int_{u_{b}(t-T_{1})}^{u_{m}(t)}\frac{du}{v_{c}}\label{eq:  tempo di propagazione andata gauge locale}\end{equation}

and

\begin{equation}
T_{2}(t)=\int_{u_{m}(t-T_{2})}^{u_{b}(t)}\frac{(-du)}{v_{c}}.\label{eq:  tempo di propagazione ritorno gauge locale}\end{equation}

The calculations of these integrals would be complicated because the
$u_{m}$ boundaries of them are changing with time:

\begin{equation}
u_{b}(t)=0\label{eq: variazione b.s. in gauge locale}\end{equation}

and

\begin{equation}
u_{m}(t)=L+\delta u_{m}(t).\label{eq: variazione specchio nin gauge locale}\end{equation}

But, to first order in $h_{+}$, these contributions can be approximated
by $\delta L_{1}(t)$ and $\delta L_{2}(t)$ (see eqs. (\ref{eq: corretto spostamento B.S. e M.})
and (\ref{eq: corretto spostamento B.S. e M. 2})). Thus, the combined
effect of the varying boundaries is given by $\delta_{1}T(t)$ in
eq. (\ref{eq: variazione tempo proprio 1}). Then, only the times
for photon propagation between the fixed boundaries $0$ and $L$
have to be computed. Such a propagation times will be indicated with
$\Delta T_{1,2}$ to distinguish from $T_{1,2}$. In the forward trip,
the propagation time between the fixed limits is

\begin{equation}
\Delta T_{1}(t)=\int_{0}^{L}\frac{du}{v_{c}(t',u)}\approx L-\int_{0}^{L}V(t',u)du,\label{eq:  tempo di propagazione andata  in gauge locale}\end{equation}

where $t'$ represents the delay time which corresponds to the unperturbed
photon trajectory (i.e. $t$ is the time at which the photon arrives
in the position $L$, so $L-u=t-t'$): 

\begin{center}
$t'=t-(L-u)$. 
\par\end{center}

Similarly, the propagation time in the return trip is

\begin{equation}
\Delta T_{2}(t)=L-\int_{L}^{0}V(t',u)du,\label{eq:  tempo di propagazione andata  in gauge locale 2}\end{equation}

where now the delay time is given by

\begin{center}
$t'=t-u$.
\par\end{center}

The sum of $\Delta T_{1}(t-T)$ and $\Delta T_{2}(t)$ gives the round-trip
time for photons travelling between the fixed boundaries. Then, the
deviation of this round-trip time (distance) from its unperturbed
value $2T$ is\begin{equation}
\begin{array}{c}
\delta_{2}T(t)=-\int_{0}^{L}[V(t-2L+u,u)du+\\
\\-\int_{L}^{0}V(t-u,u)]du,\end{array}\label{eq: variazione tempo proprio 2}\end{equation}

and, using eq. (\ref{eq:potenziale in gauge Lorentziana}), 

\begin{equation}
\begin{array}{c}
\delta_{2}T(t)=\frac{1}{2}LA\int_{0}^{L}[\int_{0}^{u}\ddot{h}_{+}(t-2T+l(1+\sin\theta\cos\phi))dl+\\
\\-\int_{0}^{u}\ddot{h}_{+}(t-l(1-\sin\theta\cos\phi)dl]du.\end{array}\label{eq: variazione tempo proprio 2 rispetto h}\end{equation}

Thus, the total round-trip proper time in presence of the GW is:

\begin{equation}
T_{t}=2T+\delta_{1}T+\delta_{2}T,\label{eq: round-trip  totale in gauge locale}\end{equation}

and\begin{equation}
\delta T_{u}=T_{t}-2T=\delta_{1}T+\delta_{2}T\label{eq:variaz round-trip totale in gauge locale}\end{equation}

is the total variation of the proper time for the round-trip of the
photon in presence of the GW in the $u$ direction.

Using eqs. (\ref{eq: variazione tempo proprio 1}), (\ref{eq: variazione tempo proprio 2 rispetto h})
and the Fourier transform of $h_{+}$, defined by

\begin{equation}
\tilde{h}_{+}(\omega)=\int_{-\infty}^{\infty}dt\textrm{ }h(t)\exp(i\omega t),\label{eq: trasformata di fourier1}\end{equation}
the quantity (\ref{eq:variaz round-trip totale in gauge locale})
can be computed in the frequency domain as 

\begin{equation}
\tilde{\delta}T_{u}(\omega)=\tilde{\delta}_{1}T(\omega)+\tilde{\delta}_{2}T(\omega)\label{eq:variaz round-trip totale in gauge locale 2}\end{equation}

where

\begin{equation}
\tilde{\delta}_{1}T(\omega)=\exp[i\omega L(1-\sin\theta\cos\phi)]LA\tilde{h}_{+}(\omega)\label{eq: dt 1 omega}\end{equation}

\begin{equation}
\begin{array}{c}
\tilde{\delta}_{2}T(\omega)=-\frac{LA}{2}[\frac{-1+\exp[i\omega L(1-\sin\theta\cos\phi)]-iL\omega(1-\sin\theta\cos\phi)}{(1-\sin\theta\cos\phi)^{2}}+\\
\\+\frac{\exp(2i\omega L)(1-\exp[i\omega L(-1-\sin\theta\cos\phi)]+iL\omega-(1-\sin\theta\cos\phi)}{(+-\sin\theta\cos\phi)^{2}}]\tilde{h}_{+}(\omega).\end{array}\label{eq: dt 2 omega}\end{equation}

In the above computation, derivative and translation Fourier transform
theorems have been used. 

Then, using eqs. (\ref{eq: A}), (\ref{eq: dt 1 omega}), (\ref{eq: dt 2 omega})
and the definition (\ref{eq: fefinizione Hu}) a \textit{signal} can
be defined:

\begin{equation}
\frac{\tilde{\delta}T_{u}(\omega)}{T}\equiv\Upsilon_{u}^{+}(\omega)\tilde{h}_{+}(\omega),\label{eq: segnale}\end{equation}

where

\begin{equation}
\Upsilon_{u}^{+}(\omega)\equiv\frac{(\cos^{2}\theta\cos^{2}\phi-\sin^{2}\phi)}{2L}\tilde{H}_{u}(\omega,\theta,\phi)\label{eq: risposta + lungo u GL}\end{equation}
 is the response function of the $u$ arm of the interferometer to
the GW.

Note: the fact that this response function is, in general, different
from zero implies that the contribution (\ref{eq: dt 1 omega}) to
the total signal, due to the motion of the test masses, will be, in
general, different from the contribution (\ref{eq: dt 2 omega}) due
to the gravitational redshift of the GW. In this way the misconception
on interferometers is clarified in the full angular and frequency
dependence of GWs. 

The same analysis works the $v$ arm. One gets the total response
function in the $v$ direction for the GWs, which is \begin{equation}
\Upsilon_{v}^{+}(\omega)\equiv\frac{(\cos^{2}\theta\sin^{2}\phi-\cos^{2}\phi)}{2L}\tilde{H}_{v}(\omega,\theta,\phi).\label{eq: risposta + lungo v GL}\end{equation}

Thus, in the gauge of the local observer, the total frequency-dependent
response function (i.e. the detector pattern) of an interferometer
to the $+$ polarization of the GW is given by:

\begin{equation}
\begin{array}{c}
\tilde{H}^{+}(\omega)\equiv\Upsilon_{u}^{+}(\omega)-\Upsilon_{v}^{+}(\omega)=\\
\\=\frac{(\cos^{2}\theta\cos^{2}\phi-\sin^{2}\phi)}{2L}\tilde{H}_{u}(\omega,\theta,\phi)+\\
\\-\frac{(\cos^{2}\theta\sin^{2}\phi-\cos^{2}\phi)}{2L}\tilde{H}_{v}(\omega,\theta,\phi),\end{array}\label{eq: risposta totale Virgo + GL}\end{equation}

which is the same result of the TT gauge (eq. (\ref{eq: risposta totale Virgo +})).
This gauge-invariance agrees with lots of results which are well known
in the literature, where the analysis has been made in the low frequency
approximation, i.e. in the case in which the wavelength of the GW
is much larger than the length of the interferometer's arms (see \cite{key-16}
for example). Note that the gauge-invariance obtained with the equality
between equation (\ref{eq: risposta totale Virgo +}) and equation
(\ref{eq: risposta totale Virgo + GL}) is more general than the one
in \cite{key-9}, where the computation was performed in the simplest
geometry of the interferometer in respect to the propagating gravitational
wave, i.e. in the case in which the arms of the interferometer are
perpendicular to the propagating GW, an only for one arm. Putting
$\theta=\phi=0$ and $v=0$ in equations (\ref{eq: risposta totale Virgo +})
and (\ref{eq: risposta totale Virgo + GL}), the result of \cite{key-9}
for one arm of an interferometer is recovered. Once again, we recall
that we do not have any idea concerning the direction of the propagating
GW that will arrive to detectors, exactly like we do not have any
idea concerning its frequency. Thus, our generalization which will
take into account both of the full angular and frequency in the gauge
of the local observer dependences is important.

A similar analysis works for the $\times$ polarization. One obtains
the same result of eq. (\ref{eq: risposta totale Virgo per}) in the
TT gauge.

Then, the total response functions of interferometers for the $+$
and $\times$ polarization of GWs, in their full angular and frequency
dependences,  are equal in the TT gauge and in the gauge of a local
observer. In this way, the gauge-invariance has been totally generalized.

\subsubsection*{Conclusions}

In this letter, the gauge-invariance on the response of interferometers
to GWs has been shown. In this process, after resuming, for completeness,
results on the Transverse-Traceless (TT) gauge, where, in general,
the theoretical computations on GWs are performed, the gauge of the
local observer has been analysed. The gauge-invariance between the
two gauges has been shown in its full angular and frequency dependences
while in previous works in the literature this gauge-invariance was
shown only in the low frequencies approximation or in the simplest
geometry of the interferometer in respect to the propagating gravitational
wave. We remark that one has not any idea concerning the direction
of the propagating GW that will arrive to detectors, exactly like
we do not know its frequency. Thus, the present results, which analyse
both of the full angular and frequency dependences in the gauge of
the local observer, are important for a sake of completeness.

As far as the computation of the response functions in the gauge of
the local observer has been concerned, a common misconception about
interferometers has been also clarified. Such a misconception purports
that, as the wavelength of laser light and the length of an interferometer's
arm are both stretched by a GW, no effect should be visible, invoking
an analogy with cosmological redshift in an expanding universe. This
issue has been raised in some papers in the literature \cite{key-17}-\cite{key-20}
and has been ultimately clarified with a new direct calculation and
without any approximation in this letter.

\subsubsection*{Acknowledgements}

The R. M. Santilli Foundation has to be thanked for partially supporting
this letter.

\end{document}